\documentclass[aps,prx,preprint,superscriptaddress]{revtex4-2}
\usepackage{latexsym}
\usepackage{graphicx}
\pdfoutput=1
\begin{document}


\title{The thickness dependence of quantum oscillations in ferromagnetic Weyl metal SrRuO$_{3}$}


\affiliation{Institute of Physics, Academia Sinica, Nankang, Taipei 11529, Taiwan}
\affiliation{High Field Magnet Laboratory (HFML-EMFL), Radboud University, Toernooiveld 7, 6525 ED Nijmegen, The Netherlands}
\affiliation{Department of Physics, Applied Physics and Astronomy, Binghamton University, Binghamton, New York 13902, USA}
\affiliation{Scientific Research Division, National Synchrotron Radiation Research Center, Hsinchu 30076, Taiwan}
\affiliation{Nano Science and Technology, Taiwan International Graduate Program, Academia Sinica and National Taiwan university.}
\affiliation{Radboud University, Institute for Molecules and Materials, Heyendaalseweg 135, 6525 AJ Nijmegen, The Netherlands}

\affiliation{Department of Physics, National Taiwan University, Taipei 10617, Taiwan}
\author{Uddipta Kar}\thanks{These authors contributed equally to the work.}
\affiliation{Institute of Physics, Academia Sinica, Nankang, Taipei 11529, Taiwan}
\affiliation{Nano Science and Technology, Taiwan International Graduate Program, Academia Sinica and National Taiwan university.}
\author{Akhilesh Kr. Singh}\thanks{These authors contributed equally to the work.}
\affiliation{Institute of Physics, Academia Sinica, Nankang, Taipei 11529, Taiwan}
\author{Yu-Te Hsu}
\affiliation{High Field Magnet Laboratory (HFML-EMFL), Radboud University, Toernooiveld 7, 6525 ED Nijmegen, The Netherlands}
\affiliation{Radboud University, Institute for Molecules and Materials, Heyendaalseweg 135, 6525 AJ Nijmegen, The Netherlands}
\author{Chih-Yu Lin}
\affiliation{Institute of Physics, Academia Sinica, Nankang, Taipei 11529, Taiwan}
\affiliation{Department of Physics, National Taiwan University, Taipei 10617, Taiwan}
\author{Bipul Das}
\affiliation{Institute of Physics, Academia Sinica, Nankang, Taipei 11529, Taiwan}
\author{Cheng-Tung Cheng}
\affiliation{Institute of Physics, Academia Sinica, Nankang, Taipei 11529, Taiwan}
\author{M. Berben}
\affiliation{High Field Magnet Laboratory (HFML-EMFL), Radboud University, Toernooiveld 7, 6525 ED Nijmegen, The Netherlands}
\affiliation{Radboud University, Institute for Molecules and Materials, Heyendaalseweg 135, 6525 AJ Nijmegen, The Netherlands}
\author{Song Yang}
\affiliation{Scientific Research Division, National Synchrotron Radiation Research Center, Hsinchu 30076, Taiwan}
\author{Chun-Yen Lin}
\affiliation{Scientific Research Division, National Synchrotron Radiation Research Center, Hsinchu 30076, Taiwan}
\author{Chia-Hung Hsu}
\affiliation{Scientific Research Division, National Synchrotron Radiation Research Center, Hsinchu 30076, Taiwan}
\author{S. Wiedmann}\email{Steffen.Wiedmann@ru.nl}
\affiliation{High Field Magnet Laboratory (HFML-EMFL), Radboud University, Toernooiveld 7, 6525 ED Nijmegen, The Netherlands}
\affiliation{Radboud University, Institute for Molecules and Materials, Heyendaalseweg 135, 6525 AJ Nijmegen, The Netherlands}
\author{Wei-Cheng Lee}\email{wlee@binghamton.edu}
\affiliation{Department of Physics, Applied Physics and Astronomy, Binghamton University, Binghamton, New York 13902, USA}
\author{Wei-Li Lee}\email{wlee@phys.sinica.edu.tw}
\affiliation{Institute of Physics, Academia Sinica, Nankang, Taipei 11529, Taiwan}


\date{\today}

\begin{abstract}
Quantum oscillations in resistivity and magnetization at high magnetic fields are a macroscopic fingerprint of the energy quantization due to the cyclotron motion of quasiparticles. In a thin Weyl semimetal, a unique thickness dependent Weyl-orbit quantum oscillation was proposed to exist, originating from a nonlocal cyclotron orbit via the electron tunneling between the top and bottom Fermi-arc surface states. Here, untwinned and high crystalline Weyl metal SrRuO$_3$ thin films with different thicknesses were grown on miscut SrTiO$_3$ (001) substrates. Magneto-transport measurements were carried out in magnetic fields up to 35 T, and quantum oscillations with different frequencies were observed and compared to the calculated band structure. In particular, we discovered a frequency $F \approx$ 30 T at low temperatures and above 3 T that corresponds to a small Fermi pocket with a light effective mass. Its oscillation amplitude appears to be at maximum for film thicknesses in a range of 10 to 20 nm, and the phase of the oscillation exhibits a systematic change with the film thickness. After isolating the well separated frequencies, the constructed Landau fan diagram shows an unusual concave downward curvature in the 1/$\mu_0H_n$-$n$ curve, where $n$ is the Landau level index. Based on the rigorous analysis of the thickness and field-orientation dependence of the quantum oscillations, the oscillation with $F \approx$ 30 T is attributed to be of surface origin, which is related to the Fermi-arc surface state originating from non-overlapping Weyl nodes projected on the film's surface plane. Those findings can be understood within the framework of the Weyl-orbit quantum oscillation effect with non-adiabatic corrections.
			
\end{abstract}


\maketitle

\newpage
\section{Introduction}
The phenomena of quantum oscillations with magnetic field in matters were first discovered in bismuth more than 90 years ago \cite{Shoenberg}, and it was soon realized that such phenomena can be used as a measure for the Fermi surface in momentum space \cite{Shoenberg,LK}. Quantitatively, quantum oscillatory phenomena are described by the general expression of $\Delta\rho \propto D $cos[$2\pi(\frac{F}{\mu_0H} + \it{\delta}$)], where $D$ is the damping factor, $F$ is the oscillation frequency, and $\it{\delta}$ is the phase shift of the oscillation associated with individual extremal areas of the Fermi surface in momentum space. In theory, the absolute value of $\it{\delta}$ is a consequence of the underlying band structure and dimensionality of the material under study in which the Berry phase is derived as 0 or $\pi$. From a modern point of view, the periodic oscillation of either $\rho$ or magnetization with respect to the inverse of a magnetic field is a quantum effect deriving from the Bohr-Sommerfeld quantization rule on electrons' cyclotron motions, where the requirement of single-valued wave function gives rise to discrete energies for electrons \cite{Onsager}. The oscillation frequency ($F$) can be linked directly to the extremal cross-sectional area ($A_{\rm e}$) of the Fermi surface by Onsager's relation of $F$ = $\hbar A_{\rm e}/2\pi e$. In addition, the amplitude and phase of quantum oscillations encode the information about the electronic band topology at Fermi surface \cite{LK} that has been a major subject recently in the field of topological materials \cite{Hasan,Armitage}.  

In topological Dirac and Weyl semimetals, a nontrivial conical band with linear dispersion results in a $\pi$ Berry phase that is picked up during the cyclotron motion and thus causes a phase shift by half a period in the quantum oscillation, which has been reported by several quantum oscillation experiments \cite{Li,Hassinger}. One practical approach for the determination of $\it{\delta}$ is to construct the Landau fan diagram, which plots the 1/$\mu_0H$ values for peaks and valleys in the quantum oscillations  (1/$\mu_0H_n$)  as a function of the corresponding Landau level (LL) index $n$. $\it{\delta}$ can then be determined from the intercept ($n_0$) as 1/$\mu_0H_n$ approaches zero by extrapolating the curve of 1/$\mu_0H_n$ versus $n$. As pointed out by several authors \cite{Ong,Shen}, the peaks (valleys) in the $\rho$ (conductivity) oscillations should be assigned with integer number of $n$ for a three-dimensional (3D) system, and thus $\it{\delta}$ = $n_0 \approx \pm$1/8 and $\pm$5/8 for a conical band with $\pi$ Berry phase and a trivial parabolic band with zero Berry phase, respectively.    

On the other hand, Weyl semimetals (WSMs) are also known for their unusual Fermi-arc surface states \cite{Arc1,Arc2} that connect the projected pair of Weyl nodes on a surface. An unusual type of quantum oscillations due to the so-called Weyl-orbit effect (WOE) was first theoretically predicted \cite{WO} and soon reported by experiments in Dirac semimetal Cd$_{3}$As$_{2}$ \cite{WOexp1,WOexp2,WOexp3}. The WOE results from an intriguing collaboration between the Fermi-arc surface states and the Weyl-node bulk states, where a non-local cyclotron orbit gives rise to a new type of quantum oscillations. One important consequence for WOE is the thickness dependence on the phase of the quantum oscillations, but, unfortunately, such a phase shift can be easily smeared out by a small density variation in different samples \cite{noWO}. It is, therefore, important to grow high crystalline thin films of WSM, and the WOE can then be carefully investigated on WSM films with different thicknesses. In this respect, the ferromagnetic oxide of SrRuO$_{3}$ (SRO) \cite{SRO1,burkov} appears to be a promising candidate. The nonmonotonic temperature ($T$) dependence of both anomalous Hall conductivity \cite{Fang} and spin wave gap in SRO \cite{Itoh} suggest the presence of the Weyl nodes near the Fermi surface that coexists with other 3D bulk Fermi pockets, making SRO a candidate for the Weyl metal phase. In addition, quantum oscillations of $\rho$ in SRO thin films on SrTiO$_{3}$ (STO) (001) substrate also revealed an unusual $F \approx$ 30 T \cite{SRO4,SRO5} that is equivalent to a small Fermi pocket with a bulk density of only 9.3 $\times$ $10^{17}$ cm$^{-3}$ for the 3D case. Such a small Fermi pocket seems to be in line with the Weyl metal phase in SRO, but its origin remains an open question. In this work, a series of highly crystalline and untwinned SRO films with different thicknesses ($t$s) ranging from about 7.7 to 35.3 nm were grown on miscut STO (001) substrates using adsorption-controlled growth by an oxide molecular beam epitaxy (MBE) system \cite{Nair2018,SROgrowth}, where pronounced quantum oscillations in $\rho$ were observed. From detailed angle dependent quantum oscillations with applied fields up to 35 T, we found that the small Fermi-pocket of $F \approx$ 30 T behaved as a two-dimensional (2D) like Fermi pocket, and the corresponding Landau fan diagram showed a clear concave downward curvature for the curve of 1/$\mu_0H_n$ versus $n$. Those results are discussed and compared to the simulated thickness dependent quantum oscillations based on WOE.        

\section{Experimental Results}
\begin{figure}
\includegraphics[width=\linewidth]{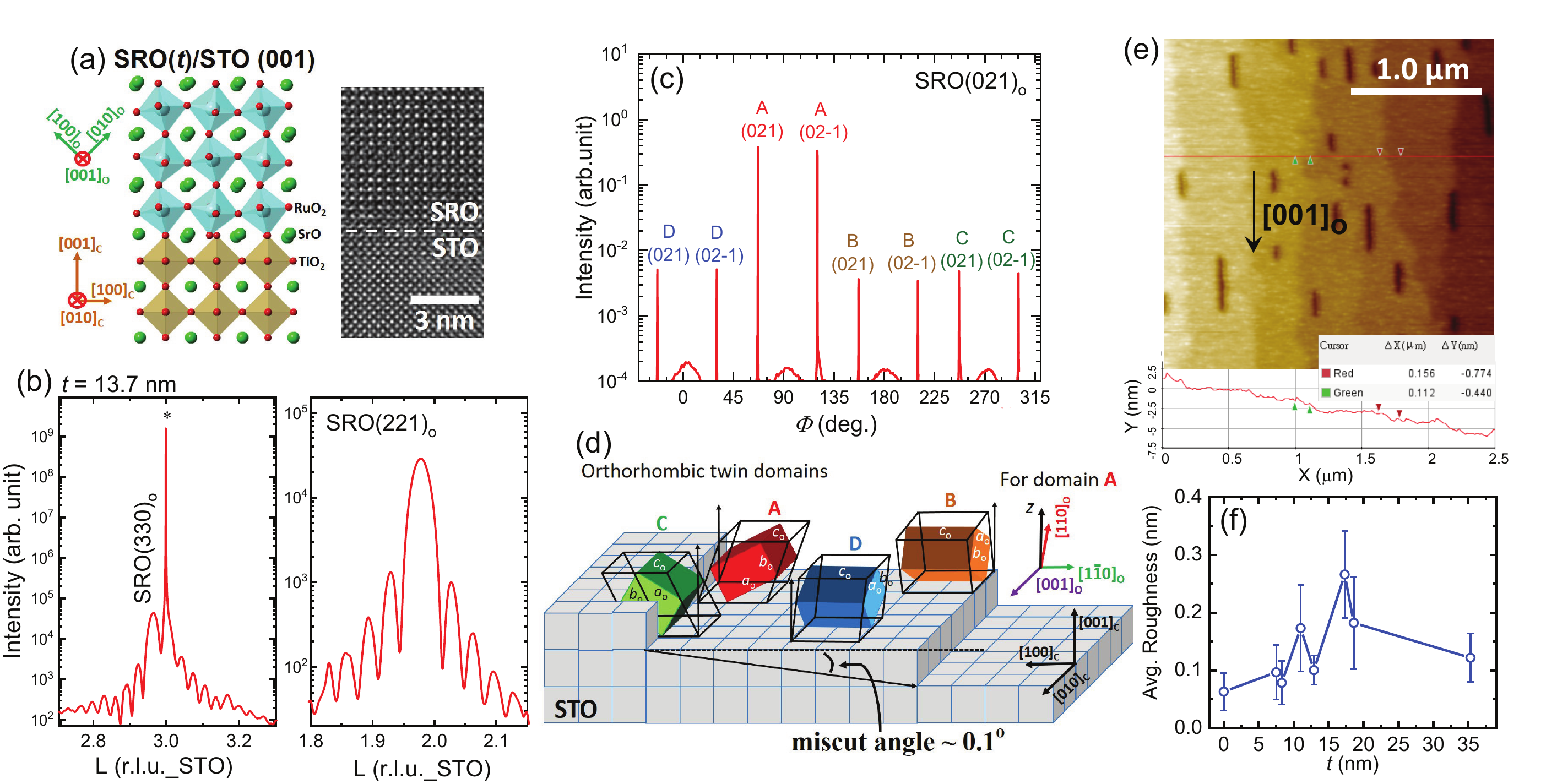}
  \caption{Structural and surface characterizations of an untwinned SRO film with $t \approx$ 13.7 nm using a miscut STO (001) substrate. (a) An atomic packing illustration (left) and the cross-sectional STEM HAADF image (right) of (110)$_{\rm o}$ oriented orthorhombic SRO on a STO (001) substrate. The left-panel and right panel of (b) show the CTRs of SRO specular (330)$_{\rm o}$ and off-normal (221)$_{\rm o}$ reflections, respectively. Pronounced fringes persisting over many periods along both CTRs manifest the excellent crystallinity and sharp interfaces of the SRO film. (c) The azimuthal $\phi$ scan across SRO (021)$_{\rm o}$ reflection reveals the presence of four orthorhombic twin domains rotated 90$^{\rm o}$ from each other as demonstrated in (d) (not to scale). The dominant domain A, amounting to 94.1 $\%$ of volume fraction, has its [001]$_{\rm o}$ direction parallel to the atomic step edges on the STO surface as verified by XRD results as well as the AFM image in (e), where the lower-panel of (e) shows the flattened cross-sectional height profile along the red line. The c-axis of SRO lies on the surface and is aligned with one of the STO lateral crystalline axis, [010]$_{\rm c}$ in the present figure. (f) The thickness dependence of the averaged surface roughness for SRO films from AFM images analyses. The averaged roughness is less than 0.3 nm, which does not vary significantly for $t \leq$ 35.3 nm. 
  }
  \label{CTR}
\end{figure}

\begin{figure}
\includegraphics[width=\linewidth]{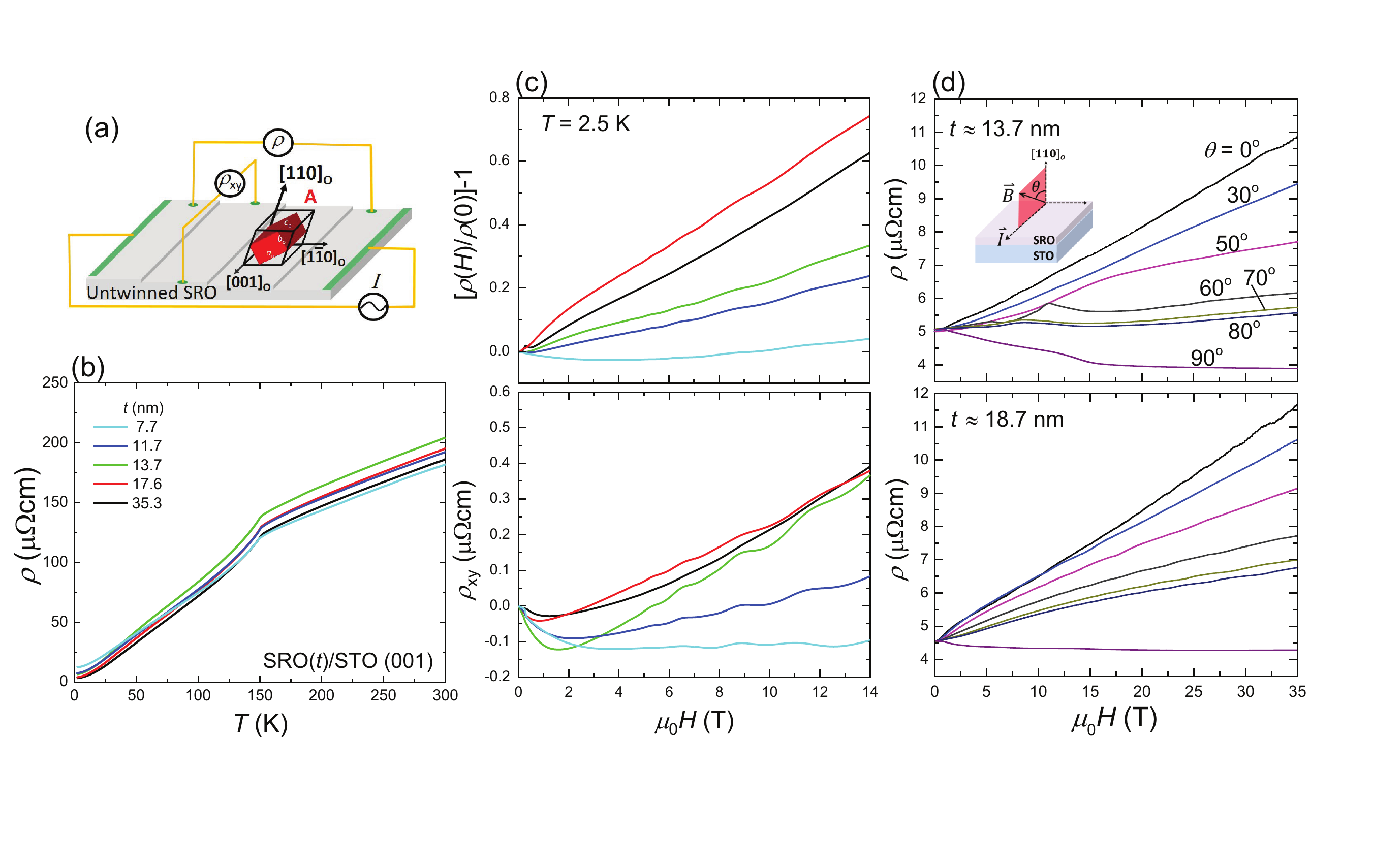}
  \caption{The temperature and field dependence of $\rho$ and $\rho_{\rm xy}$ in SRO thin films. (a) an illustration of the measurement geometry for $\rho$ and $\rho_{\rm xy}$ with respect to the SRO crystalline axes. (b) The $T$ dependent $\rho$ in SRO($t$)/STO (001) with thicknesses $t$s ranging from about 7.7 to 35.3 nm. The kink at $T \approx$ 150 K corresponds to the onset of the ferromagnetism. The upper-panel and lower-panel of (c) show the MR and $\rho_{\rm xy}$, respectively, as a function of field at $T$ = 2.5 K, and they share the same color codes as that in (b). Pronounced quantum oscillations were observed for $\mu_{0}H \geq$ 3.5 T in all five SRO films with different $t$ values. (d) The $\theta$ dependent MR for $t \approx$ 13.7 nm and 18.7 nm are shown in the upper-panel and lower-panel, respectively. The inset cartoon shows the definition of the $\theta$ angle. For field perpendicular to the film surface plane ($\theta$ = 0$^{\rm o}$), a large and positive MR was observed, which is practically linear with the field strength up to 35 T without saturation. The magnitude of the positive MR drops with increasing $\theta$, and it becomes a negative MR when the field is applied along the current direction ($\theta $ = 90$^{\rm o}$). }
  \label{rt}
\end{figure}

The crystal structure of perovskite SRO is illustrated in the left-panel of Fig. \ref{CTR}(a), where the rotation and tilting of RuO$_{6}$ octahedra give rise to an orthorhombic phase at room temperature. When growing on a STO (001) substrate, a compressive strain of about $-$0.4 $\%$ in SRO films occurs and gives rise to a distorted orthorhombic phase with the crystalline axis of SRO [110]$_{\rm o}$ nearly aligned along the STO [001]$_{\rm c}$, where the subscripts o and c refer to the orthorhombic- and cubic-phase, respectively \cite{SRO1}. The right-panel of Fig. \ref{CTR}(a) shows a cross-sectional scanning transmission electron microscope (STEM) high-angle annular dark-field (HAADF) image of a SRO film we grew, indicating an atomically sharp interface between SRO and STO. High resolution X-ray measurements were performed, and the left-panel of Fig. \ref{CTR}(b) displays the STO (003)$_{\rm c}$ crystal truncation rods (CTRs) for a SRO film with $t\approx$ 13.7 nm. The abscissa L is in unit of STO reciprocal lattice unit (r.l.u.), representing the momentum transfer along the surface normal. The SRO (330)$_{\rm o}$ reflection appeared at slight low-L side of STO (003)$_{\rm c}$ reflection \cite{Nair2018}. The presence of pronounced thickness fringes around the SRO Bragg reflection indicates excellent film crystallinity and sharp interfaces. The oscillation period of about 0.031 r.l.u. gives a crystalline thickness of about 12.6 nm, which is close to $t \approx$ 13.7 nm as obtained from the X-ray low angle reflectivity measurements. The right-panel of Fig. \ref{CTR}(b) shows the CTR of SRO (221)$_{\rm o}$ off-normal reflection. The observation of the orthorhombic specific SRO (221)$_{\rm o}$ reflection with pronounced intensity oscillations confirmed the orthorhombic phase and also excellent thickness uniformity and crystallinity of the SRO film \cite{Chang2011}. The corresponding azimuthal $\phi$ scan across the SRO (021)$_{\rm o}$ reflection is shown in Fig. \ref{CTR}(b), where two sets of equally spaced peaks with different peak intensity were observed. This indicates the presence of four orthorhombic twin domains \cite{Gan1997} as illustrated in Fig. \ref{CTR}(d). Be aware that the [110]$_{\rm o}$ direction of orthorhombic SRO can be considered as parallel to the normal of (110)$_{\rm o}$ plane because of the tiny ($<$ 0.8 $\%$) difference between the length of its a- and b-axis that was highly exaggerated in Fig. \ref{CTR}(d). From the peak intensity, the volume fractions for domain A, B, C and D are estimated to be about 94.1, 1.7, 2.0 and 2.2 $\%$, respectively, which thus justifies the nearly single-domain and untwinned structure in our SRO films. Detailed investigations of the orthorhombic twin domains for SRO films grown on miscut STO (001) substrates have been reported in a separate paper \cite{SROgrowth}. On the other hand, the surface morphology of the SRO film with $t \approx$ 13.7 nm was measured by an atomic force microscope (AFM) as shown in Fig. \ref{CTR}(e), where the lower-panel shows the cross-section height profile across the red-line located in the upper-panel of Fig. \ref{CTR}(e). The step heights at the terrace edges are about 1 to 2 unit cells of STO, which again supports for the high thickness uniformity of the SRO film. However, we do note the presence of well isolated pits that is known to exist for the step-flow growth of SRO film \cite{SRO1}. By excluding those pits, the averaged roughness of the SRO film can be obtained from the AFM images, and the resulting $t$-dependence of the averaged roughness is shown in Fig. \ref{rt}(f). For $t \leq$  35.3 nm, the average roughness of the film was determined to be less than about 0.3 nm (see supplementary section 1). We also note that the dominating domain is found to have SRO [001]$_{\rm o}$ (the black arrow in Fig. \ref{CTR}(e)) along the terrace edge on a STO substrate. For the growth of SRO films in this work, all the STO substrates were chosen to have a small miscut angle of about 0.1$^{\rm o}$ with the terrace edge being parallel to a lateral STO $<$100$>$ direction to ensure the single-domain in our SRO films \cite{SROgrowth}. In addition, the bias current for the transport measurements was set to be perpendicular to the terrace edge as demonstrated in Fig. \ref{rt}(a), i.e. $\vec{I} \perp$ SRO [001]$_{\rm o}$, in order to avoid complication and inconsistency due to possible $\rho$ anisotropy and domain dependence in SRO.

The $T$-dependent $\rho$ exhibits a metallic behavior as shown in Fig. \ref{rt}(b) for SRO films with different $t$s ranging from about 7.7 to 35.3 nm. The kink at $T \approx$ 150 K indicates the occurrence of a ferromagnetic transition, which does not vary significantly with $t$ within the range of our interests. At lower temperatures, $\rho$ follows a $T^2$ dependence as predicted for a Fermi liquid system \cite{SRO4}, and $\rho$ at $T$ = 5 K ($\rho$(5 K)) shows progressive increase from $\rho$(5 K) $\approx$ 3.8 to 12.7 $\mu\Omega$cm as the SRO film thickness decreases from $t \approx$ 35.3 to 7.7 nm. $\rho$ in ruthenates is known to show high sensitivity to disorder \cite{rhoSRO}, and thus the low $\rho$(5 K) values indicate the high crystallinity of our SRO films. A positive transverse magnetoresistance (MR) with $H \parallel$ SRO [110]$_{\rm o}$ was observed for thicker SRO films with $t \geq$ 10 nm as shown in the upper-panel of Fig. \ref{rt}(c) for $T$ = 2.5 K, and the MR, defined as [$\rho(H)/\rho(H=0)$] $-$ 1, can be as large as +74 $\%$ at $\mu_0H$ = 14 T for SRO film with $t \approx$ 17.6 nm. It gradually decreases with decreasing $t$, giving a MR of about +24 \% at $\mu_0H$ = 14 T for SRO film with $t \approx$ 11.7 nm.  For $t \approx$ 7.7 nm, the MR first decreases with increasing field to a value of $-$3 $\%$ at  $\mu_0H \approx$ 3.6 T above which it increases with field instead, giving a small positive MR $\approx$ +4 \% at $\mu_0H$ = 14 T. On the other hand, the field-dependent Hall resistivity ($\rho_{\rm xy}$) shows a sign change in the Hall slope (d$\rho_{\rm xy}$/d$\mu_0H$) from negative to positive as $\mu_0H$ goes from 0 to 14 T, which was plotted in the lower-panel of Fig. \ref{rt}(c) for $T$ = 2.5 K. Pronounced quantum oscillations were clearly observed in both $\rho$ and $\rho_{\rm xy}$ for $\mu_0H \geq$ 3 T, which is only possible for highly crystalline SRO films with low residual resistivity \cite{SRO4,SRO5}. When tilting $H$ toward the applied current direction ($\vec{I}$) by an angle $\theta$ as illustrated in the inset cartoon of Fig. \ref{rt}(d), the MR changes from being positive to negative as $\theta$ goes from 0$^{\rm o}$ to 90$^{\rm o}$, as shown in the upper-panel and lower-panel of Fig. \ref{rt}(d) for SRO films with $t \approx$ 13.7 nm and 18.7 nm, respectively, at $T \approx$ 0.3 K. Remarkably, the transverse MR ($\theta$ = 0$^{\rm o}$) is practically $H$-linear without any tendency of saturation, giving a positive MR of about +158 $\%$ and +117 $\%$ for $t \approx$ 13.7 nm and 18.7 nm, respectively, at $\mu_0H$ = 35 T. On the other hand, for $\theta$ = 90$^{\rm o}$ with $H \parallel$ $\vec{I}$, apparent negative MRs were observed in the low field regime, giving a negative MR of about $-$20 $\%$ and $-$5 $\%$ at $\mu_0H \sim$ 15 T for $t \approx$ 13.7 nm and 18.7 nm, respectively. The $H$-linear transverse MR and the negative longitudinal MR are reminiscent of the MR behaviors commonly observed in topological Dirac and Weyl semimetals \cite{Armitage,Xiong,Zhang,SRO5}.

\begin{figure}
\includegraphics[width=\linewidth]{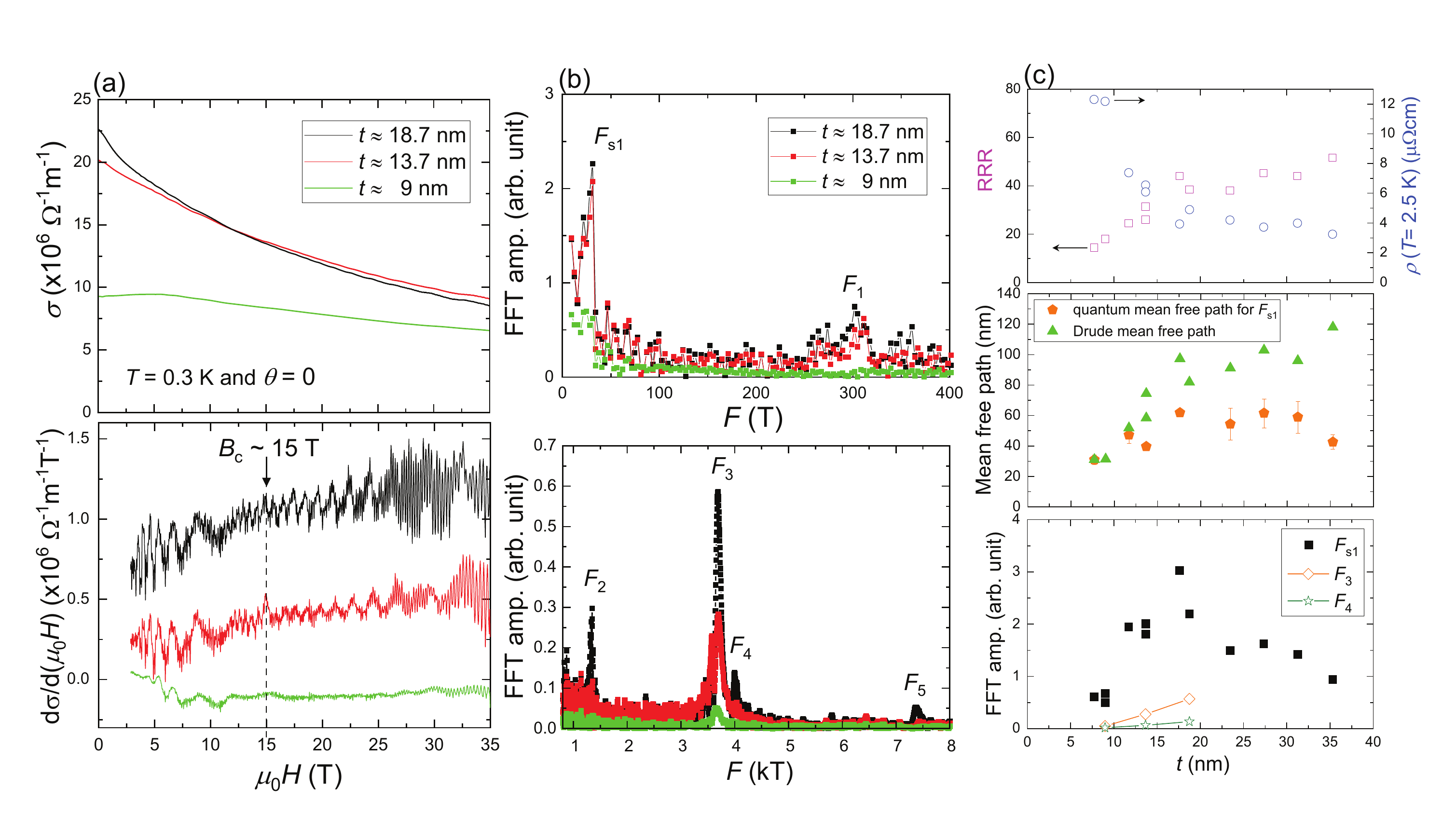}
  \caption{The thickness dependence of quantum oscillations in SRO films. The upper-panel of (a) shows the field dependent $\sigma$ with fields up to 35 T for three SRO films with different $t$ values, and the corresponding d$\sigma$/d($\mu_0H)$ versus $\mu_0H$ were plotted in the lower-panel, where pronounced quantum oscillations were observed for $\mu_0H \geq$ 3.5 T. The upper-panel and lower-panel of (b) shows the FFT spectra in low-frequency region and high-frequency region, respectively. The quantum oscillation for $F_{\rm s1}$ (about 30 T) dominates in the low field region, and it seems to vanish for $\mu_0H \geq$ 15 T. Other frequencies of $F_1$ to $F_5$ appear when $\mu_0H \geq$ 12 T. A summary of the thickness dependence of extracted mean free path and FFT amplitudes are shown in the middle-panel and lower-panel, respectively. The nonmonotonic variation of the FFT amplitude for $F_{\rm s1}$ with $t$ is in big contrast to the nearly $t$-linear dependence for $F_3$ and $F_4$. The corresponding RRR and $\rho$(2.5 K) as a function of $t$ were plotted in the upper-panel of (c). }
  \label{HH}
\end{figure}

The resulting field dependent conductivity ($\sigma \equiv \rho_{\rm}/(\rho^2_{\rm} + \rho^2_{\rm xy})$) for the three different $t$s ranging from 9.0 to 18.7 nm at $T$ = 0.3 K is shown in the upper-panel of Fig. \ref{HH}(a). In zero field, $\sigma$ equals about 9.3 $\times$ 10$^6$ $\Omega^{-1}$m$^{-1}$ for $t \approx$ 9.0 nm, and it increases with $t$, giving a $\sigma \approx$ 22.6 $\times$ 10$^6$ $\Omega^{-1}$m$^{-1}$ for $t \approx$ 18.7 nm. $\sigma$ drops with increasing fields up to 35 T, and the corresponding d$\sigma$/d($H$) versus $\mu_0H$ was plotted in the lower-panel of Fig. \ref{HH}(a), where pronounced quantum oscillations were observed for $\mu_0H \geq$ 3.5 T. By taking the fast Fourier transform (FFT) on the curves of d$\sigma$/d$H$ versus 1/$\mu_0H$, several distinct frequencies of $F_{\rm s1}$ and $F_{1-5}$ can be clearly identified from the FFT spectra as shown in Fig. \ref{HH}(b). The oscillations with $F_{1-5}$ frequencies ranging from $\approx$ 300 to 7,400 T appeared to show up when $\mu_0H \geq$ 12 T with the oscillation amplitudes grow with increasing film thickness $t$. On the other hand, the oscillation for $F_{\rm s1} \approx$ 30 T starts at a much lower field of $\approx$ 3.5 T, and it appears to vanish when the field is above a critical value of $B_{\rm c} \sim$ 15 T. Following the standard practice, the effective mass ($m^*$) and Dingle temperature ($T_{\rm D}$) can be determined from the damping of the quantum oscillations \cite{Shoenberg}. From the temperature dependent measurements, the effective mass was determined to be $m^*$ = (0.30 $\pm$ 0.01) $m_{\rm e}$ for $F_{\rm s1}$, where $m_{\rm e}$ is the electron mass. On the other hand, the extracted $m^*$ values for the Fermi pockets with $F \approx$  300 and 3,680 T are more than tenfold larger with $m^*$ = (3.2 $\pm$ 0.3) and (5.5 $\pm$ 0.8) $m_{\rm e}$, respectively (see supplementary section 2). A summary of the extracted parameters from quantum oscillation data is shown in Table \ref{tab}. $\tau_{\rm q} \equiv \hbar/2\pi k_{\rm B}T_{\rm D}$ is the quantum lifetime. The resulting $\ell_{\rm q} \equiv (\hbar k_{\rm F}/m^*)\tau_{\rm q}$ is the quantum mean free path, and the Drude mean free path $\ell_{\rm d}$ was calculated using a bulk density of about 6.1 $\times$ 10$^{21}$ cm$^{-3}$ for all $t$s. At $T$ = 2.5 K, $\ell_{\rm q}$ falls in a range from 31 to 62 nm with a relatively weak dependence on $t$ as demonstrated in the middle-panel of Fig. \ref{HH}(c). On the contrary, $\ell_{\rm d}$ at $T$ = 2.5 K monotonically increases by nearly four-folds from 31 to 118 nm as $t$ increases from 7.7 to 35.3 nm.

\begin{table}
\caption{A summary of the extracted $F$, $m^*$, $T_{\rm D}$, $\tau_{\rm q}$ and $\ell_{\rm q}$ from the quantum oscillation measurements at $T$ = 2.5 K, where $\ell_{\rm d}$ values are included for comparison. The maximum field ($B_{\rm max}$) and $T$-range of the measurement are also listed.} 
\centering 
\begin{tabular}{|c|c|c|c|c|c|c|c|} 
\toprule
$t$ (nm) &{\it{F} $\rm (T)$} & $m^*$ ($m_{\rm e}$) & $T_{\rm D}$ (K) & $\tau_{\rm q}$ (ps) &$\ell_{\rm q}$ (nm)& $\ell_{\rm d}$ (nm) & $B_{\rm max}$ (T)/$T$-range (K)\\ 
\colrule
7.7   & $F_{\rm s1} \approx$ 30 & 0.37 & 3.71 &  0.33 & 31& 31 & 14/2.5--10\\
11.7 & $F_{\rm s1} \approx$ 30 & 0.33 &2.72 & 0.45 & 47 & 52 & 14/2.5--10\\
13.7 & $F_{\rm s1} \approx$ 30 & 0.31 & 3.53 & 0.34 & 40& 58 & 14/2.5--10\\
17.6 &$F_{\rm s1} \approx$ 30 & 0.30 & 2.36 & 0.51 & 62 & 97 & 14/2.5--10\\
35.3 & $F_{\rm s1} \approx$ 30 & 0.25 & 4.00 & 0.30 &43 & 118& 14/2.5--10\\
\colrule
9.0 & $F_{\rm s1} \approx$ 30 & 0.30 &1.94 &0.64 &69& 31 & 35/0.3--4.2 \\
13.7 &$F_{\rm s1} \approx$ 30 & 0.32 & 2.41 & 0.51 &53 & 75 & 35/0.3--4.2 \\
18.7 & $F_{\rm s1} \approx$ 30  & 0.27 &2.18 & 0.55 & 69& 82 & 35/0.3--4.2 \\
18.7 & $F_{1} \approx $  300 & 3.20 & 3.25 & 0.37 &13 &-& 35/0.3--4.2 \\ 
18.7 & $F_{3} \approx$ 3,680 & 5.50 & 5.84 & 0.21 &15 & -&35/0.3--4.2 \\
\botrule 
\end{tabular}
\label{tab}
\end{table}

Figure \ref{HH}(c) summarizes the thickness dependence of residual resistivity ratio (RRR $\equiv \rho$(300 K)/$\rho$(5 K)) (upper-panel), mean free path (middle-panel), and FFT amplitudes (lower-panel). As $t$ increases from 7.7 to 35.3 nm, the RRR follows an increasing trend from 14.4 to 51.6, and the corresponding $\rho$(2.5 K) decreases from 12.3 to 3.2 $\mu\Omega$cm. The observed values of $F_{\rm s1}$,  $F_{\rm s2}$, $F_3$ and $F_4$ show relatively weak $t$-dependence, but the FFT amplitude for $F_{\rm s1}$ appears to show nonmonotonic $t$ dependence. In particular, for 10 $\leq t \leq$ 20 nm, the FFT amplitude for $F_{\rm s1}$ is maximal with a weak $t$ dependence, regardless of a nearly 50 $\%$ variation in RRR as shown in the upper-panel of Fig. \ref{HH}(c). This is in sharp contrast to the nearly $t$ linear dependence of FFT amplitudes for $F_{3}$ and $F_{4}$ as expected for the increased RRR with a larger $t$. Such a nonmonotonic $t$ dependence on the FFT amplitude indicates that the quantum oscillation for $F_{\rm s1}$ is more likely to be surface origin.  

\section{Electronic band calculations}
\begin{figure}
\includegraphics[width=\linewidth]{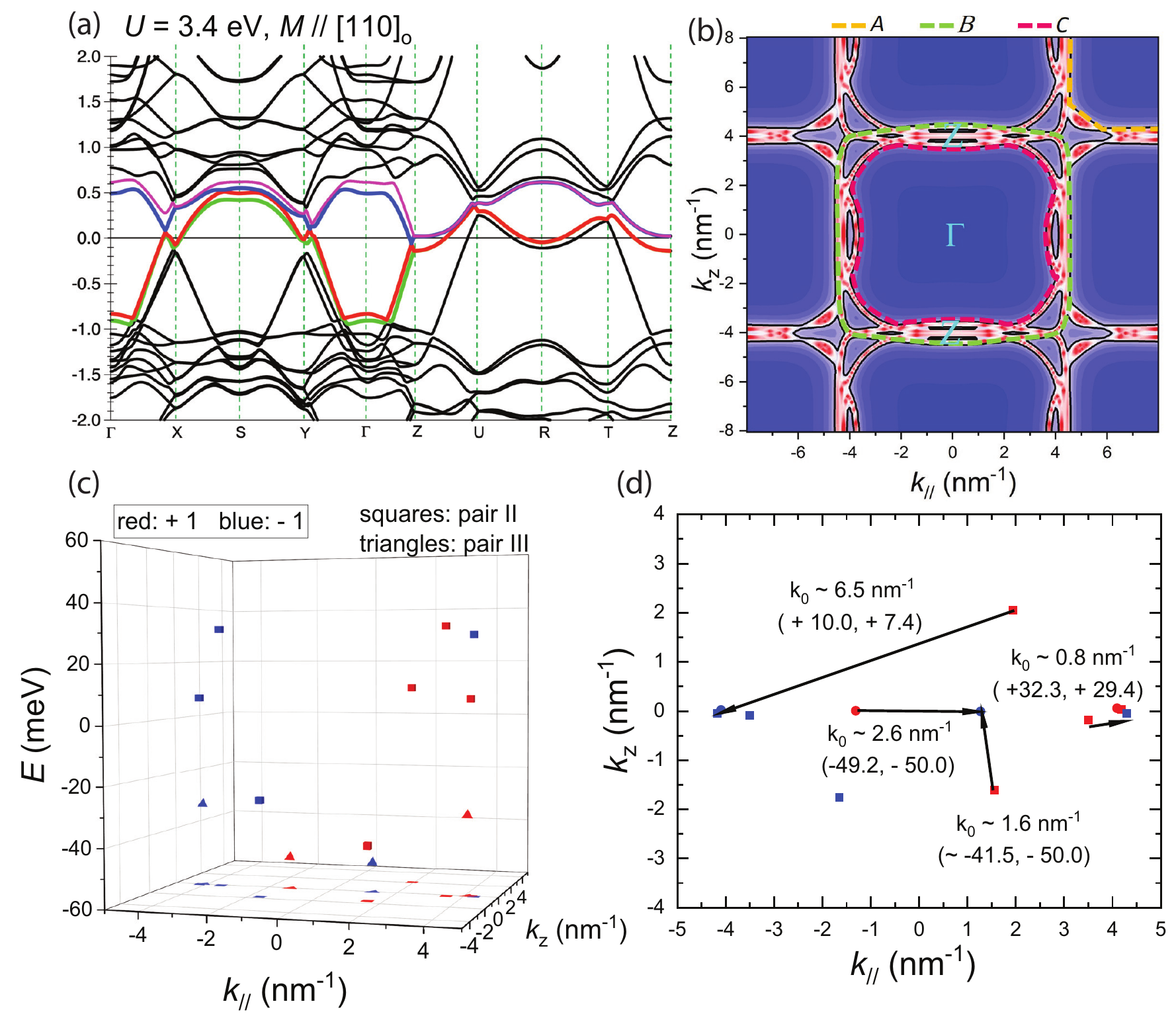}
\caption{ Electronic structures calculated by DFT + U with $U = 3.4$ eV and $J = 0.68$ eV for the orthorhombic SRO. (a) Band structure near the Fermi energy calculated with spin-orbit coupling in the ferromagnetic ground state with the magnetic moment oriented along [110]$_{\rm o}$ direction. We searched for the Weyl nodes in four pairs of bands near the Fermi energy (pair I $\to$ green band and the one below; pair II $\to$ (green, red); pair III $\to$ (red, blue); pair IV $\to$ (blue, pink) ) (b) The corresponding FS plot sliced along the $k_\parallel$-$k_{\rm z}$ plane, which indicates three major pockets of A, B and C. (c) Locations of Weyl nodes residing in these four pairs in the energy window of $\vert E-E_F\vert < 50$ meV. The red and blue symbols are Weyl nodes with a chirality of +1 and $-$1, respectively. (d) The non-overlapping Weyl-node pairs projected on $k_\parallel$-$k_{\rm z}$ plane. The arrows connect from +1 to $-$1 Weyl nodes at similar energy, where the calculated $k_{\rm 0}$ values and the corresponding Weyl node energies in unit of meV are indicated.  }
	\label{band}
\end{figure}

We have calculated electronic structures using first-principles methods \cite{wien2k,wien2k2020,anisimov1993,wannier90,wanniertools}. For $U = 3.4$ eV and $J = 0.68$ eV, we obtain a ferromagnetic (FM) ground state with the magnetic moment on Ru sites close to 1.20 $\mu_B$, which agrees well with our experiments and previous DFT study \cite{jeng2006}. Moreover, we find that the ground state energy of the system with the magnetic moment oriented along [110]$_{\rm o}$ direction is slightly lower than that along [001]$_{\rm o}$ direction ($E^{FM}_{110} - E^{FM}_{001} \approx -0.546$ meV/f.u.), which is consistent with the observed magnetic easy axis along [110]$_{\rm o}$ in our SRO films. The calculated band structures are shown in Fig. \ref{band}(a), and a number of Weyl nodes were clearly identified when searching in four subband pairs near the Fermi energy (see supplementary section 3). In order to identify the Fermi pockets for the observed quantum oscillations shown in Fig. \ref{HH}, the $k_{\rm x}$ and $k_{\rm y}$ in the original momentum space were transformed into $k_{\parallel}$ and $k_{\perp}$ with $k_{\parallel}$ lying in the (110)$_{\rm o}$ plane. The calculated Fermi surface (FS) sliced along $k_\parallel$-$k_{\rm z}$ plane with $k_{\perp}$ = 0 is shown in Fig. \ref{band}(b) with the center being the $\Gamma$ point. We identified three major pockets of A, B and C with areas of 48, 72 and 53 nm$^{-2}$, respectively. The corresponding frequencies are 5077, 7531 and 5517 T, which are relatively close to the observed $F_{3} \approx$ 3680 T, $F_{5} \approx$ 7400 T and $F_{4} \approx$ 4000 T, respectively, as shown in the lower-panel of Fig. \ref{HH}(b). The A and B pockets are well consistent with the observed FS from an earlier angle-resolved photo-emission experiment \cite{ARPES2013}. We notice several smaller pockets between the A and B, but, unlike the three major pockets, they changed dramatically as $k_{\perp}$ varies across the $\Gamma$ point (see supplementary section 4). Therefore, those smaller pockets may not contribute coherently to give observable quantum oscillations. We further moved on to look for possible non-overlapping Weyl nodes when projecting on (110)$_{\rm o}$ plane. The energy dependence of the Weyl nodes at ($k_\parallel$, $k_{\rm z}$) within a window of $\vert E - E_F\vert \leq 50$ meV is shown in Figure \ref{band}(c), where the squares and triangles are Weyl nodes from subband pair II and subband pair III, respectively, with the red (blue) color being the chirality of +1 ($-$1). A number of non-overlapping Weyl nodes projected on $k_\parallel$-$k_{\rm z}$ plane are clearly observed. By selecting the projected Weyl nodes at similar energies with opposite chiralities, four Weyl-node pairs are identified and shown in Fig. \ref{band}(d), where the black arrows represent wave vectors connecting Weyl-node pairs from +1 to $-$1. The corresponding Fermi-arc lengths fall in a range from $k_0 \approx$ 0.8 to 6.5 nm$^{-1}$. Those band calculation results support for the presence of Fermi-arc surface states on (110)$_{\rm o}$ plane, which is essential for the occurrence of WOE. However, we remark that, due to the complex band crossings near the Fermi energy, the Weyl nodes locations are highly sensitive to several parameters used for the band calculations. Nevertheless, non-overlapping Weyl-node pairs projected on (110)$_{\rm o}$ plane are always present for different $U$ values ranging from 2 to 3.4 eV (see supplementary section 3).   
             
\section{Quantum Oscillations analyses}
\begin{figure}
\includegraphics[width=\linewidth]{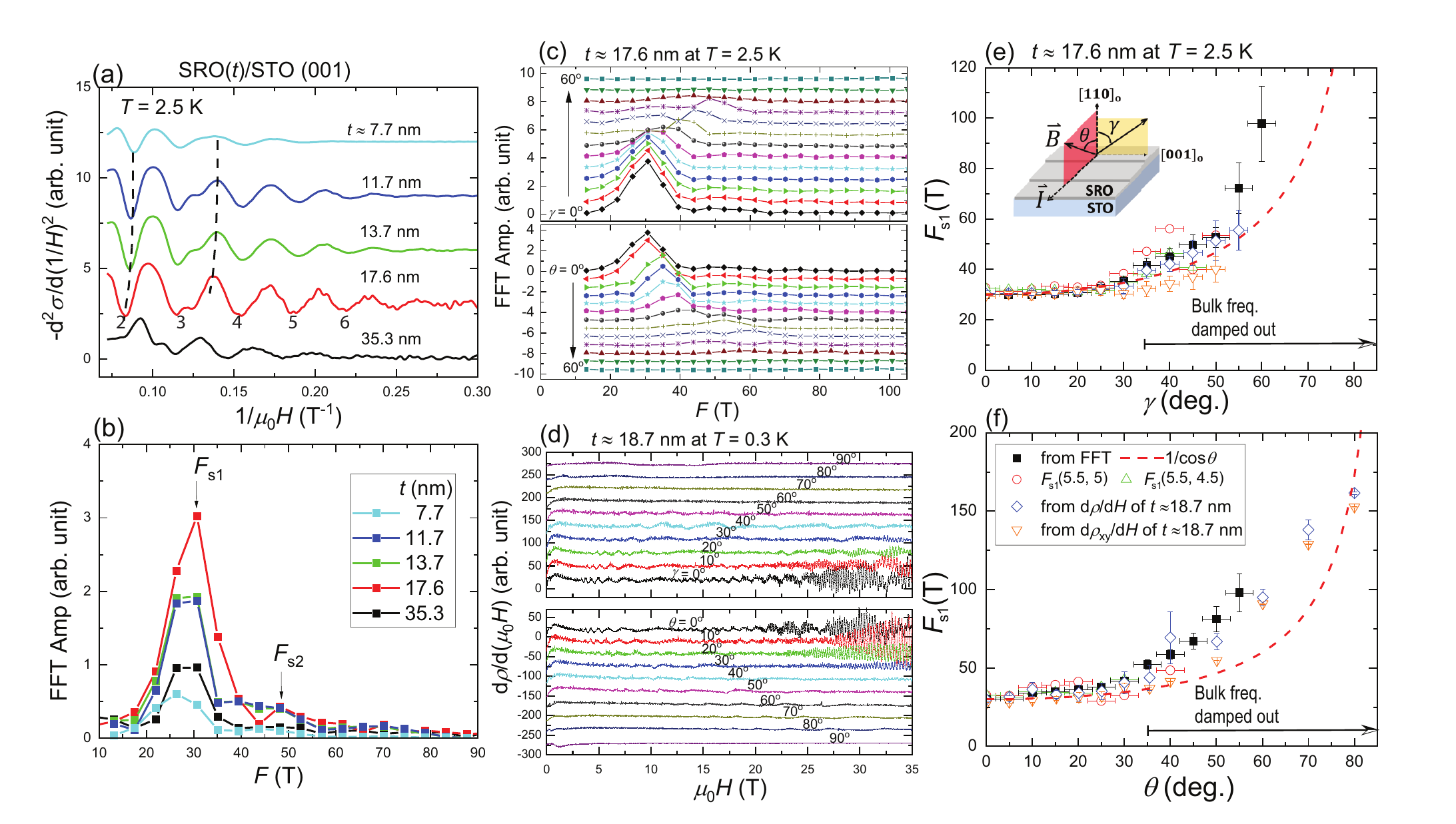}
  \caption{The quantum oscillations of SRO films in the low field regime. (a) The $-$d$^2\sigma$/d(1/$H)^2$ versus 1/$\mu_0H$ at $T$ = 2.5 K for SRO films with five different $t$s ranging from about 7.7 to 35.3 nm. The curves with different $t$s are shifted vertically for clarity. The corresponding FFT spectra are shown in (b). The oscillation and FFT amplitude for $F_{\rm s1}$ appear to show larger magnitude for $t$s in the range of about 10 to 20 nm. For $t \approx$ 17.6 nm, the angle dependent FFT spectra for different $\gamma$ and $\theta$ angles are shown in the upper-panel and lower-panel, respectively, in (c). For  $t \approx$ 18.7 nm, the angle dependent quantum oscillations of d$\rho$/d($\mu_0H)$ versus $\mu_0H$ up to 35 T were plotted in (d), where the bulk frequencies of $F_{1-5}$ damped out at angles larger than 30$^{\rm o}$. The curves with different $\theta$ and $\gamma$ values are shifted vertically for clarity. The extracted $F_{\rm s1}$ as a function of $\gamma$ and $\theta$ angles are shown in (e) and (f), respectively. The red dashed-lines represent the 1/cos$\theta$ and 1/cos$\gamma$ dependence. The inset cartoon in (e) illustrates the orientation for the $\gamma$ and $\theta$ angles.     
    }
  \label{SdH}
\end{figure}

For better revelations of the quantum oscillations in $\sigma$ without subtracting any artificial background, we plot $-$d$^2\sigma$/d(1/$\mu_0H)^2$ as a function of 1/$\mu_0H$ as shown in Fig. \ref{SdH}(a) for SRO films with different $t$s ranging from about 7.7 to 35.3 nm. The peak and valley locations in the oscillation turn out to show systematic phase shifts to lower 1/$\mu_0H$ values as $t$ increases, indicating by long dashed lines in Fig. \ref{SdH}(a), and an integer number of LL index ($n$) was assigned to the minimum of $-$d$^2\sigma$/d(1/$H)^2$ \cite{Armitage,Ong}. The corresponding FFT spectra for SRO films with different $t$ values are shown in Fig. \ref{SdH}(b). A dominant frequency of $F_{\rm s1} \approx$ 30 T was clearly observed, and another frequency of $F_{\rm s2} \approx$ 50 T was also observable with a much smaller FFT amplitude and also broader distribution in the FFT spectra. Therefore, the quantum oscillations for $\mu_0H \leq$ 14 T as shown in Fig. \ref{SdH}(a) are dominated by an unusual small Fermi pocket associated with $F_{\rm s1}$, which turns out to show the largest magnitude for $t \approx$ 17.6 nm. A decrease of the FFT amplitude occurs for $t >$ 20 nm and $t <$ 10 nm as is evident from the results in Fig. \ref{SdH}(a). In order to further probe the Fermi pocket associated with $F_{\rm s1}$, detailed angle dependence of the quantum oscillations were performed with $\gamma$-rotation and $\theta$-rotation setups as illustrated in the inset of Fig. \ref{SdH}(e), where the direction of magnetic field ($\vec B$) was tilted toward the current ($\vec{I}$) and SRO [001]$_{\rm o}$ for $\theta$-rotation and $\gamma$-rotation, respectively. $\theta$ and $\gamma$ values are the angles between $\vec B$ and the surface normal of SRO films, i.e. SRO [110]$_{\rm o}$ that appears to be the magnetic easy axis with the lowest coercive field (see supplementary section 5). The resulting FFT spectra of $\gamma$-rotation and $\theta$-rotation for SRO film with $t \approx$ 17.6 nm and fields up to 14 T at $T =$ 2.5 K are shown in the upper-panel and lower-panel, respectively, of Fig. \ref{SdH}(c) for angles ranging from 0$^{\rm o}$ to 60$^{\rm o}$. As the angle increases from 0$^{\rm o}$ to 60$^{\rm o}$, $F_{\rm s1}$ gradually shifts to higher frequencies for both $\theta$-rotation and $\gamma$-rotation. Figure \ref{SdH}(d) shows the $\theta$ and $\gamma$ dependent quantum oscillations of d$\rho$/d($\mu_0H)$ versus $\mu_0H$ for SRO thin film with $t \approx$ 18.7 nm and fields up to 35 T at $T =$ 0.3 K. For angles larger than about 30$^{\rm o}$, the bulk frequencies of $F_{\rm 1-5}$ damp out rapidly as expected in thin films with increased surface scatterings. The extracted $F_{\rm s1}$ as a function of $\gamma$ and $\theta$ are summarized in Fig. \ref{SdH} (e) and (f), respectively, where $F_{\rm s1}$(5.5, 5) and $F_{\rm s1}$(5.5, 4.5) are extracted from the oscillation periods of (1/$\mu_0H_{5.5}$ $-$ 1/$\mu_0H_5$) and (1/$\mu_0H_{5.5}$ $-$ 1/$\mu_0H_{4.5}$), respectively. We included the results for the SRO film with $t \approx$ 18.7 nm shown as diamond and downward-triangle symbols in Fig. \ref{SdH}(e-f), which was measured with applied fields up to 35 T. However, we do note that, at some angles of around 30$^{\rm o}$, $F_{\rm s1}$ in the FFT spectra seems to split into two frequencies but then merged into a single peak again at higher angles. Nevertheless, the extracted $F_{\rm s1}$ values for all SRO films with different $t$s tend to follow the 1/cos$\theta$ or 1/cos$\gamma$ dependence (red-dashed lines in Fig. \ref{SdH} (e-f)), indicating a 2D-like nature of the Fermi pocket for $F_{\rm s1}$ (see supplementary section 5). 

By taking the oscillation period of adjacent peaks and valleys in Fig. \ref{SdH}(a) for $\theta$ = 0$^{\rm o}$, the extracted $F_{\rm s1}$ as a function of the LL index $n$ are shown in the upper panel of Fig. \ref{Weyl}(a) for SRO films with $t \approx$ 13.7 and 17.6 nm. Here, $n$ was determined by setting $B_{\rm c}$ as the field above which the quantum limit occurs. $F_{\rm s1}$ marches down from about 34.5 T for $n$ = 8.5 to 28 T for $n$ = 4, and this behavior is consistent with the observed downward bending of the 1/$\mu_0H_n$-$n$ curve in the Landau fan diagram shown in the lower-panel of Fig. \ref{Weyl}(a). Surprisingly, the intercept $n_0$, obtained by linearly extrapolating the high $n$ data (black-dashed line), gives an unusual large phase shift of $\it{\delta} \approx -$2.0.

\section{Discussions}
\begin{figure}
\includegraphics[width=\linewidth]{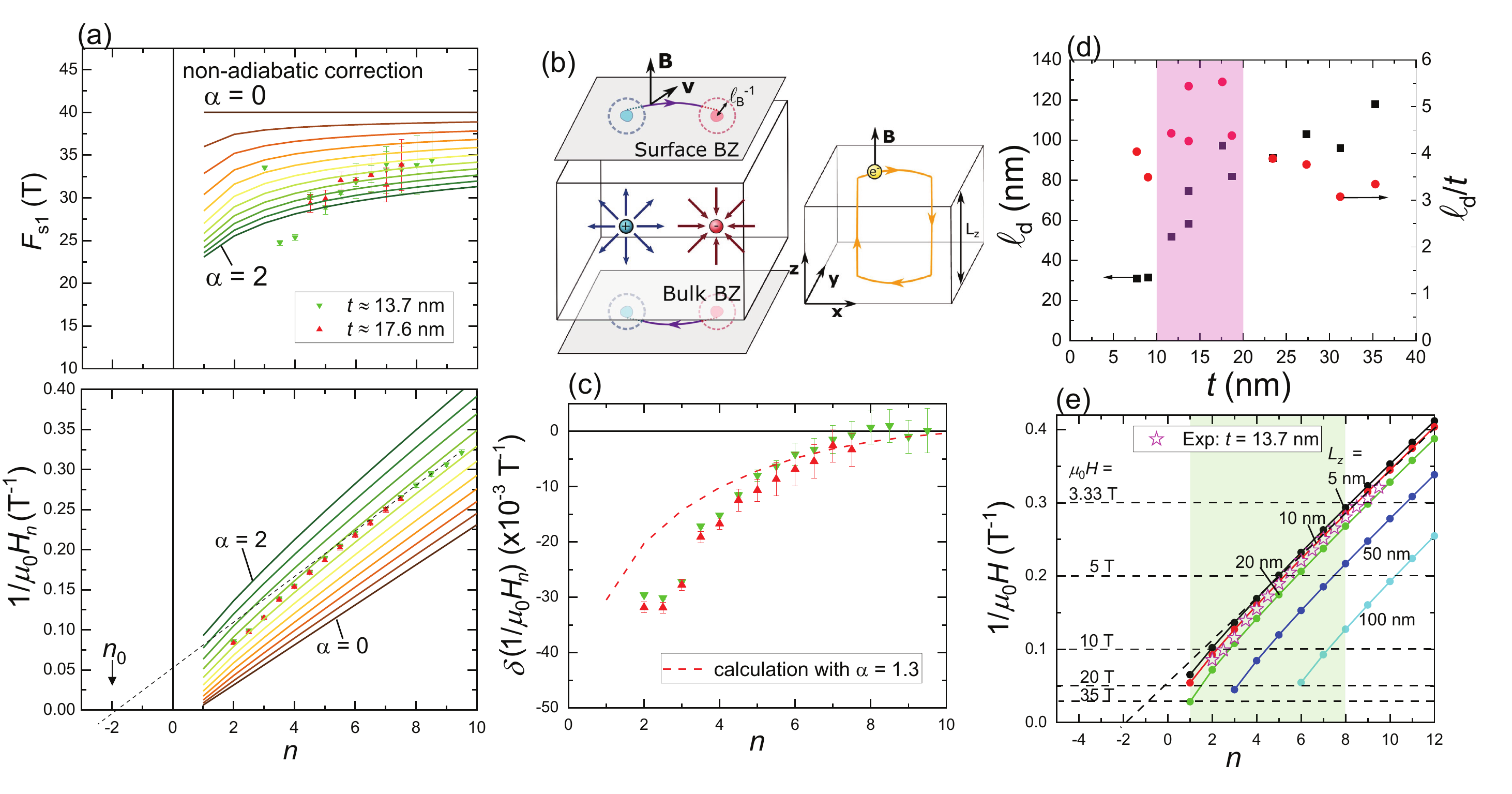}
  \caption{ The comparison to simulated Weyl orbit quantum oscillations. (a) The simulated curves based on the WOE formula with non-adiabatic correction are plotted and compared to the experimental data from SRO thin films with $t \approx$ 13.7 and 17.6 nm, where the color codes for the solid lines represent the $\alpha$ parameters ranging from 0 to 2.0. The experimental data are found to fall close to the simulated curves with $\alpha \approx$ 1.3, exhibiting both a unique downward bending curvature in 1/$\mu_0H_n$-$n$ curve and also an unusual large phase shift $n_0 \approx - 2$. (b) An illustration of the nonlocal cyclotron orbit resulting from WOE. (c) shows the deviation from linearity $\delta$(1/$\mu_0H_n$) as a function of $n$ due to non-adiabatic correction, which is in good agreement with the simulated curve with $\alpha$ = 1.3. (d) The extracted Drude mean free path ($\ell_{\rm d}$) and ratio of $\ell_{\rm d}/t$ as a function of $t$, showing a largest $\ell_{\rm d}/t$ for $ 10 < t < 20$ nm. (e) The simulated 1/$\mu_0H_n$-$n$ curves for $\alpha$ = 1.3 with different $t$s based on WOE, inferring an optimum thickness of about $t \sim$ 10 to 20 nm for the dominant WOE in SRO thin films.}
  \label{Weyl}
\end{figure}

The surface nature of $F_{\rm s1} \approx$ 30 T with a small effective mass $m^*$ = (0.30 $\pm$ 0.01) $m_{\rm e}$ was strongly supported by the 2D-like angle dependence of $F_{\rm s1}$ and nonmonotonic $t$ dependence of oscillation amplitude as shown in Fig. \ref{SdH}(e-f) and Fig. \ref{HH}(c), respectively. Considering a conventional 2D parabolic band with a frequency of 30 T, the field required for the quantum limit ($B_{\rm q}$) can be calculated, giving $B_{\rm q} \approx$ 30 T and 15 T for spin-degenerate bands and single spin subband, respectively. We also remark that such a 2D Fermi pocket is not likely coming from the SRO/STO interface conduction due to possible band bending or oxygen loss from STO, which typically gives a larger effective mass of $m^* >$ 1.0 $m_{\rm e}$ \cite{STOmass1,STOmass2}. One intriguing possibility for such a 2D-like small pocket of $F_{\rm s1} \approx$ 30 T is then the unusual quantum oscillation due to WOE as illustrated in Fig. \ref{Weyl}(b), which is supported by the existence of nonoverlapped Weyl-node pairs near the Fermi energy from our band calculations (Fig. \ref{band}) and also several earlier reports \cite{SRO5,burkov}. The dashed circles around the projected Weyl nodes with a length scale of inverse magnetic length ($\ell_{\rm B}^{-1}$) are boundaries where the transition from the Fermi-arc surface states to the bulk states happens before reaching the ends of the Fermi-arcs, giving rise to the reduced effective area of a Weyl-orbit and thus the non-adiabatic correction effect \cite{WO}. The quantum oscillation due to WOE can be described as \cite{WO} 
\begin{equation}
\frac{1}{\mu_0H_{n}} = \frac{e}{\hbar k_0'}[\frac{\pi\hbar\upsilon n}{\mu} - t],
\label{WOeq}
\end{equation}
where $n$ is the LL index, $\mu$ is the chemical potential, and $\upsilon$ is the Fermi velocity. $ k_0' \equiv k_0(1 - 4\alpha/k_0\ell_{\rm B})$ is defined as the effective wave vector connecting the projected Weyl-node pair on the surface, and $k_0$ is the length of the Fermi-arc. $\ell_{\rm B} \equiv \sqrt{\hbar/e\mu_0H_n}$ is the magnetic length, and $\alpha$ is the parameter governing the non-adiabatic correction \cite{WO}. The corresponding frequency can then be calculated using $F_{\rm s1}$ = 1/$\mu_0H_n$ - 1/$\mu_0H_{n-1}$. One important consequence of Eq. \ref{WOeq} is the thickness dependence of the phase in the oscillation, and the 1/$\mu_0H_{n}$ value will progressively decrease with increasing $t$ values and the amount of phase shift should be independent of $n$, which is in accordance with our observation shown in Fig. \ref{SdH}(a). We remark that the phase shift due to trivial minor carrier density variations is expected to be linearly proportional to $n$, which is thus strikingly different from the $n$-independent phase shift due to WOE (see supplementary section 7). By solving for 1/$\mu_0H_{n}$ in Eq. \ref{WOeq}, the resulting simulated $F_{\rm s1}$-$n$ and 1/$\mu_0H_n$-$n$ curves are shown as solid lines in the upper-panel and lower-panel of Fig. \ref{Weyl}(a), respectively, where the gradual change of line color represents the progressive increase of $\alpha$ parameter from 0 to 2. Here, we used $m^*$ = 0.3 $m_{\rm e}$, $t =$ 13.7 nm, and the $F_{\rm s1}$ = 40 T for $\alpha$ = 0.            

For $\alpha$ = 0, $F_{\rm s1}$ is independent of $n$, and the 1/$\mu_0H_n$-$n$ curve is strictly linear with an intercept of $n_0 \approx$ 0.76. When taking into account the non-adiabatic correction effect ($\alpha \neq$ 0), several notable variations occur. First, the overall $F_{\rm s1}$ values shift to lower values, and the resulting $F_{\rm s1}$-$n$ curves show concave downward feature with $F_{\rm s1}$ values decreases with decreasing $n$. Secondly, the 1/$\mu_0H_n$-$n$ curve in the Landau fan diagram also shows gradual upward-shifting with increasing concave downward curvature as $\alpha$ increases. We note that the concave downward curvature in the 1/$\mu_0H_n$-$n$ curve is distinct from the behavior due to the Zeeman energy contribution that causes a concave upward curvature instead \cite{WOexp1}. The experimental data of $F_{\rm s1}$ and 1/$\mu_0H_n$ are shown as downward-triangles and upward-triangles in Fig. \ref{Weyl}(a) for $t \approx$ 13.7 and 17.6 nm, respectively, which turns out to agree well with the simulated curves with $\alpha \approx$ 1.3. The amount of deviation $\delta$(1/$\mu_0H_n$) from linearity as a function of $n$ is shown in Fig. \ref{Weyl}(c), where the slope of the linear background derives from the 1/$\mu_0H_{n}$ curve for $\alpha$ = 0. The simulated curve of $\delta$(1/$\mu_0H_n$)-$n$ (red dashed line in Fig. \ref{Weyl}(c)) with $\alpha$ = 1.3 shows quantitative agreements with the experimental data for SRO films with $t \approx$ 13.7 and 17.6 nm. We note that the simulated curves for different $\alpha$ were calculated solely from Eq. \ref{WOeq} without imposing additional scaling parameters (see supplementary section 7). Once $\alpha$ = 1.3 is determined, the corresponding WOE parameters can then be calculated, giving $k_0$ =  1.09 nm$^{-1}$, $\mu$ = 15.43 meV, and $\upsilon$ = 1.34 $\times$ 10$^5$ ms$^{-1}$. The extracted $k_0$ value falls in the same order of magnitude as the calculated $k_0$ as shown in Fig. \ref{band}(d). We also note that there can be multiple Weyl-node pairs on (110)$_{\rm o}$ plane that support for multiple Weyl-orbits with different frequencies, such as $F_{\rm s2}$. However, due to the small oscillation amplitude for $F_{\rm s2}$, it is not possible to resolve its intrinsic mechanism with current data. In particular, we remark that the non-adiabatic correction effect, i.e. a finite $\alpha$, can, in principle, shift the 1/$\mu_0H_n$-$n$ curve in the Landau fan diagram, resulting a significant change in the intercept $n_0$. Therefore, an unusual large phase shift of $\it{\delta} \approx -$2.0 is possible for the WOE with the non-adiabatic correction, which can be another unique feature for quantum oscillations deriving from WOE. 

In general, the phase shift of the $\rho$ oscillations in a 2D system can be expressed as $\it{\delta}$ = $\frac{1}{2} + \frac{\Phi_B}{2\pi}$ (modulo 1), where $\phi_B$ is the Berry phase, and the magnitude of $\it{\delta}$ is not likely to exceed 1.0. On the other hand, the phase shift of quantum oscillations from a simple 3D bulk Weyl-node should give a $\it{\delta}$ close to $\pm$1/8 \cite{Armitage,Ong,Shen}. It was pointed out theoretically \cite{Shen} that a large anomalous phase shift can occur in a 3D topological WSM when the Fermi level is close to the Lifshitz point at which both the linear and parabolic bands are important. According to the theory \cite{Shen}, the model Hamiltonian can be expressed as ${\cal H}$ = $A$($k_{\rm x}\sigma_{\rm x}$ + $k_{\rm y}\sigma_{\rm y}$) + $M$ ($k^2_{\rm w} - k^2$), where $A$ and $M$ are the coefficients for the linear and parabolic band, respectively. $k_{\rm w}$ is the separation of the Weyl-node pair. Two energy scales of $E_{\rm A} \equiv Ak_{\rm w}/2$ and $E_{\rm M} \equiv Mk^2_{\rm w}/4$ govern the phase shift behavior, where an anomalous large phase shift appears for $E_{A} \leq E_{\rm M}$. By using the parameters from our quantum oscillation data of $F_{\rm s1}$, $E_{\rm A}$ and  $E_{\rm M}$ are estimated to be about 48.5 meV, and 2.3 meV, where we used $k_{\rm w}$ = $k_0$ and an effective mass of 5 $m_{\rm e}$ for the parabolic band. Therefore, the observed quantum oscillations in our SRO films do not fall in the condition for the large anomalous phase shift due to the coexist of the linear and parabolic bands.

In a Weyl metal, the observed quantum oscillations can derive from complex competitions among several different mechanisms, such as the WOE, 3D Weyl-nodes, and 3D bulk pockets. In our SRO thin films, the quantum oscillation of $F_{\rm s1}$ we observed for $t \leq$ 40 nm is not likely dominated by 3D Weyl-node bulk states, since its amplitude shows nonmonotonic variation with $t$ (lower-panel of Fig. \ref{HH}(c)) and also its angle dependence behaves more close to a 2D-like Fermi surface (Figs. \ref{SdH}(e-f)). On the other hand, the occurrence of quantum oscillations due to WOE requires a high uniformity of film thickness and also $t \ll \ell_{\rm d}$, which are both well satisfied in our SRO films as shown in Fig. \ref{CTR} and Table \ref{tab}. Phenomenologically, the oscillation amplitude of WOE is expected to be proportional to exp(-$t/\ell_{\rm d}$) \cite{WOexp1}. The ratio of $\ell_{d}/t$ versus $t$ in SRO thin films are well above 3 for $t <$ 40 nm and attends a maximum values for 10 $\leq t \leq$ 20 nm as shown in Fig. \ref{Weyl}(d), inferring an optimum thickness range for a maximum oscillation amplitude due to WOE. In addition, the condition of $\ell_{d}/t \geq$ 3 also supports for observed rapid damping of the bulk frequencies of $F_{\rm 1-5}$ due to the increased surface scattering for angles larger than 30$^{\rm o}$ as shown in Fig. \ref{SdH}(d). On the other hand, we also simulated the $t$ dependence of 1/$\mu_0H_n$-$n$ curves with $\alpha$ = 1.3 as demonstrated in Fig. \ref{Weyl}(e), where an unusual large phase shift of $n_0 \sim -$2 can only be observed for $t \leq$ 20 nm. As $t$ increases to 50 nm or larger, the 1/$\mu_0H_n$-$n$ curve shift downward to smaller values of 1/$\mu_0H_n$, and thus a much higher field strength is required to access the low $n$ states, posing a inevitable limitation on the lowest $n$ that can be reached by available field strength. In other words, for $t >$ 50 nm, the WOE amplitude and field accessible range in $n$ are both reduced, and the contribution from 3D bulk pockets may become dominant instead, making it difficult to identify the quantum oscillations due to WOE.                  

At last, the observed $B_{\rm c} \sim$ 15 T is not the saturation field ($B_{\rm sat} \equiv$ ($\hbar k_{\rm F}k_{0}$/$\pi e$)) for the WOE. For $\mu_0H \geq B_{\rm sat}$, the Weyl orbit mostly happens within the bulk states, and thus the associated quantum oscillations vanish \cite{WO,WOexp1}. Using the extracted parameters in our SRO films, $B_{\rm sat}$ is estimated to be about 80 T and greater than $B_{\rm c}$. However, a precise identification of the quantum limit regime for $F_{\rm s1}$ is challenging and may call for further investigations. It may require not only measurements with field strength larger than 35 T but also a precise extraction of $F_{\rm s1}$ contribution from a large oscillating background ($F_{1-5}$) that grows rapidly with field strength as shown in Fig. \ref{Weyl}(b) (see supplementary section 8).                 

\section{Conclusions}
The revelation of quantum oscillations due to WOE in Dirac and Weyl systems is challenging due to the mixing of quantum oscillations from 3D bulk Fermi pockets. Combining quantum oscillation measurements in SRO thin films and WOE simulations, we identified an optimum thickness of $t \sim$ 10-20 nm for achieving a dominant WOE contribution, where several unique features associated with WOE were clearly observed. Starting from the growth of a series of untwinned SRO films on miscut STO (001) substrates with different $t$s by using adsorption-controlled growth technique with an oxide MBE, the RRR ($\rho$(2.5 K)) shows a progressive increase (decrease) with film thicknesses ranging from 7.7 to 35.3 nm, and the condition of $t < \ell_{\rm d}$ is always satisfied to favor the bulk tunneling and thus WOE, where the measurement geometry was kept the same with respect to the SRO orthorhombic crystalline direction. The calculated band structure of SRO shows complex band crossings and supports for the existence of non-overlapping Weyl nodes when projecting on the film surface plane. The observed $H$-linear transverse MR with fields up to 35 T and also negative longitudinal MR in our SRO films further support for its topological Weyl metal phase. From rigorous angle and thickness dependent measurements of quantum oscillations with fields up to 35 T, we revealed an unusual quantum oscillation with a small frequency of $F_{\rm s1} \approx$ 30 T, which behaves like a 2D Fermi pocket with a small effective mass of about 0.3 $m_{\rm e}$. Its oscillation amplitude shows nonmonotonic $t$-dependence and attends a maximum when $t$ is in the range of 10 to 20 nm, where a systematic phase shift of the oscillation with $t$ was observed. In addition, the extracted 1/$\mu_0H_n$-$n$ curve in the Landau fan diagram shows a unique concave downward curvature. Those observations agree well with the WOE formalism with non-adiabatic corrections, corresponding to a Fermi-arc length ($k_0$) of 1.09 nm$^{-1}$ and a non-adiabatic correction parameter ($\alpha$) of about 1.3. With an excellent control on the growth for high crystalline and untwinned oxide thin films, the topological Weyl metal SRO can be an ideal platform to explore the exotic phenomena associated with the bulk Weyl nodes and Fermi-arc surface states. 

$Note$ $added.$ We are aware of a recent related quantum oscillations data in SRO films \cite{yamamoto2}, and similar 2D-like small Fermi pocket with a frequency of 30 T was reported and attributed to Fermi-arc surface states. The discussions in our work are focus on the untwinned SRO film thickness regime with $t < \ell_{\rm d}$, particularly for $t$ in a range from 10 to 20 nm, and the effect of non-adiabatic correction turns out to be an important ingredient to explain the observed features in Weyl-orbit quantum oscillations.     

\section*{Methods}
The adsorption-controlled growth of SRO thin films was carried out using an oxide MBE system. The detailed growth conditions and structural characterization was reported in an earlier paper \cite{SROgrowth}. For SRO films with $t \approx$ 9, 13.7 and 18.7 nm, the magnetotransport measurements with applied fields up to 35 T were carried out at High Field Magnet Laboratory in Nijmegen. For other $t$s and another $t \approx$ 13.7 nm of the same batch, the magnetotransport measurements were performed using a superconducting magnet with fields up to 14 T. The high resolution X-ray measurements were performed at TPS 09A and TLS 07A of the NSRRC in Taiwan, where the orthorhombic phase, film thicknesses, and orientation of the SRO films were carefully examined (see supplementary section 1). The surface roughness of SRO films and the STO substrate miscut orientation were measured by an AFM. The SRO films presented in this work were carefully selected to have the same STO miscut orientation with respect to the direction of the probing currents for the magnetotransport measurements, i.e. the current direction is perpendicular to the terrace edge of the STO (001) substrate. 

The DFT + U calculations were performed with full potential linear augmented plane waves plus local orbitals (FP-LAPW + lo) and the Perdew-Burke-Ernzerhof generalized gradient approximation (PBE-GGA) provided in the WIEN2k code.\cite{wien2k,wien2k2020} For the treatment of Hubbard $U$ terms, we have adopted the self-interaction corrections developed by Anisimov et al. \cite{anisimov1993} which is available as SIC scheme in WIEN2k package. We have used the crystal lattice constants of $a_0$ = 5.584 \AA, $b_0$ = 5.540 \AA, and $c_0$ = 7.810 \AA, which were determined by high resolution X-ray measurements on our samples, and a $k$-mesh of $22\times 22\times 15$ was used to sample the Brillouin zone. We have performed calculations with several different values of $U$ with $J$ fixed to be $J$ = 0.2 $U$. All the DFT + U calculations were done with the inclusion of spin-orbit coupling on the heavy atoms of Sr and Ru.

The tight-binding model employed for the calculation of Weyl nodes was composed of the $d$-orbitals of the Ru atoms as well as the $p$ orbitals of the O atoms, and the hopping parameters are determined by fitting the DFT band structure using the Wannier90 \cite{wannier90}. We focused on five bands near the Fermi energy, as shown in Fig. 6(a) of the main text. We searched for all the band crossings residing in four pairs of bands out of these five bands. These four pairs are defined as: pair I $\to$ green band and the one below; pair II $\to$ (green, red); pair III $\to$ (red, blue); pair IV $\to$ (blue, pink). Finally, we confirmed the Weyl nodes by calculating the chirality of each band crossing using the WannierTools \cite{wanniertools}.

The Fermi surfaces are plotted using the spectral function which is defined as
\begin{equation}
A(\vec{k},E) = \frac{1}{\pi}{\rm Im}\left\{{\rm Tr}\left[\frac{1}{E-\hat{H}_{TB}(\vec{k}) -i\epsilon}\right]\right\},
\end{equation}
where $\hat{H}_{TB}(\vec{k})$ is the tight-binding model fitted by the Wannier90, and we have used the codes provided by WannierTools \cite{wanniertools} for evaluating the spectral function.
   
\section*{Data Availability}
All the supporting data are included in the main text and also in supplementary information. The raw data and other related data for this paper can be requested from W.L.L. 

\section*{Code Availability}
The input files for DFT calculations using WIEN2k are available upon request. 

\section*{Acknowledgements}
This work was supported from Academia Sinica (Thematic Research Program), Ministry of Science and Technology of Taiwan (MOST Grant No. MOST 108-2628-M-001 -007 -MY3 and and MOST 106-2112-M-213-006-MY3) and HFML-RU/NWO-I, a member of the European Magnetic Field Laboratory (EMFL).   

\section*{Competing interests} The authors declare no competing financial or non-financial interests.

\section*{Author Contributions}
C.Y.L., B.D., C.T.C. and W.L.L. carried out the low temperature magneto-transport measurements and data analyses. U.K. and A.K.S. grew the epitaxial SRO films. Y.T.H., M.B. and S.W. performed magnetotransport measurements at HMFL in Nijmegen. U.K., A.K.S., S.Y., C.Y.L. and C.H.H. performed the X-ray measurements at NSRRC in Taiwan. W.C.L. performed SRO band calculations. W.C.L. and W.L.L. designed the experiment and wrote the manuscript.  

\section*{Additional Information}
\textbf{Supplementary Information} accompanies the paper on the XXXX website (https://XXXXX).\\



\clearpage

\newpage

\end{document}